\documentclass[%
prd,
nofootinbib,
superscriptaddress,
]{revtex4}
\usepackage{amsmath}
\usepackage{amssymb}
\usepackage{braket}
\usepackage{bm}
\usepackage{graphicx}
\begin{document}
\title{Weak lensing from self-ordering scalar fields}
%======================= Author =======================%
\author{Shohei Saga}
\affiliation{Department of Physics, Nagoya University, Aichi 464-8602, Japan}
\email{saga.shohei@nagoya-u.jp}
\author{Kouichirou Horiguchi}
\affiliation{Department of Physics, Nagoya University, Aichi 464-8602, Japan}
\author{Kiyotomo Ichiki}
\affiliation{Department of Physics, Nagoya University, Aichi 464-8602, Japan}
\affiliation{Kobayashi-Maskawa Institute for the Origin of Particles and the Universe, Nagoya University, Nagoya-city, Aichi 464-8602, Japan}
%======================= Abstract =======================%
\begin{abstract}
Cosmological defects result from cosmological phase transitions in the early Universe and the dynamics reflects their symmetry-breaking mechanisms.
These cosmological defects may be probed through weak lensing effects because they interact with ordinary matters only through the gravitational force.
In this paper, we investigate global textures by using weak lensing curl and B modes.
Non-topological textures are modeled by the non-linear sigma model (NLSM), and induce not only the scalar perturbation but also vector and tensor perturbations in the primordial plasma due to the nonlinearity in the anisotropic stress of scalar fields.
We show angular power spectra of curl and B modes from both vector and tensor modes based on the NLSM.
Furthermore, we give the analytic estimations for curl and B mode power spectra.
The amplitude of weak lensing signals depends on a combined parameter $\epsilon^{2}_{v} = N^{-1}\left( v/m_{\rm pl} \right)^{4}$ where $N$ and $v$ are the number of the scalar fields and the vacuum expectation value, respectively.
We discuss the detectability of the curl and B modes with several observation specifications.
In the case of the CMB lensing observation without including the instrumental noise, we can reach $\epsilon_{v} \approx 2.7\times 10^{-6}$.
This constraint is about 10 times stronger than the current one determined from the Planck.
For the cosmic shear observation, we find that the signal-to-noise ratio depends on the mean redshift and the observing number of galaxies as $\propto z^{0.7}_{\rm m}$ and $\propto N^{0.2}_{\rm g}$, respectively.
In the study of textures using cosmic shear observations, the mean redshift would be one of the key design parameters.
\end{abstract}
%======================= Make title =======================%
\maketitle
%============== Section ==============%
\section{introduction}
%Phase transition and defects
Current cosmological observations confirm that the universe begins with extremely high temperature, what we call the hot big-bang model.
As the universe expands adiabatically, it cools down from the hot initial condition.
Therefore, it is natural to expect that cosmological phase transitions occur in the history of the universe.
Cosmological phase transitions result in various cosmological defects depending on the symmetry of the phase transitions, e.g., cosmic strings, domain walls, and textures, which were first discussed by T.W.B.Kibble \cite{1976JPhA....9.1387K}.
We can examine the nature of the phase transition that happened in the early universe through the resulting defects by using cosmological observations since these defects affect various observables; in the case of cosmic strings, see e.g., Ref.~\cite{Hindmarsh:1994re}.

%Texture
The global $O(N)$ symmetry breaking results in domain walls ($N = 1$), cosmic strings ($N = 2$), monopoles ($N = 3$), textures ($N = 4$), and non-topological textures ($N > 4$).
Effects of the defects such as cosmic strings and textures can be seen at the horizon scale at that time, which corresponds to the correlation length of the strings or the textures.
According to this fact, defects could affect several cosmological observables in the various scales through the metric perturbations, which include, for example, gravitational waves \cite{PhysRevD.64.064008,2000PhRvL..85.3761D,2012PhRvD..86b3503K,2014PhRvD..89b3512B,2010PhRvD..81j3523K,2016JCAP...11..005M}, weak gravitational lensings \cite{2007PhRvD..76l3515M,2008MNRAS.384..161K}, generation of magnetic fields \cite{2016PTEP.2016h3E02H}, the cosmic microwave background (CMB) angular power spectrum \cite{1999PhRvD..60h3504P,2012PhRvD..86l3513A} and the CMB lensing \cite{Yamauchi:2012bc,Yamauchi:2011cu,Namikawa:2013wda}.

In this paper, we focus on the non-topological texture with large-$N$ limit $N\gg 4$ \cite{PhysRevLett.63.2625,PhysRevLett.66.3093,Kunz:1996ka}.
The dynamics of non-topological textures is exactly described by the non-linear sigma model (NLSM).
Effects of textures on the cosmological observations, such as the large-scale structure \cite{PhysRevLett.63.2625,Spergel:1990ee}, cosmic microwave background  fluctuations \cite{Durrer:1997te,Durrer:1998rw,GarciaBellido:2010if,Fenu:2013tea,Durrer:2014raa}, gravitational waves \cite{Figueroa:2012kw,Fenu:2009qf,JonesSmith:2007ne}, and generation of magnetic fields \cite{Horiguchi:2015xsa}, have been studied in many articles.
Some cosmological defects including textures induce not only the scalar, but also the vector and tensor modes originated from the anisotropic stress of scalar fields such as \cite{PhysRevD.64.064008,2000PhRvL..85.3761D,2012PhRvD..86b3503K,2014PhRvD..89b3512B,2010PhRvD..81j3523K,2016JCAP...11..005M,2007PhRvD..76l3515M,2008MNRAS.384..161K,2016PTEP.2016h3E02H,1999PhRvD..60h3504P,2012PhRvD..86l3513A,Yamauchi:2012bc,Yamauchi:2011cu,Namikawa:2013wda,Durrer:1997te,Durrer:1998rw,GarciaBellido:2010if,Fenu:2013tea,Durrer:2014raa,Figueroa:2012kw,Fenu:2009qf,JonesSmith:2007ne,Horiguchi:2015xsa}.
These vector and tensor modes are good tracers of cosmological defects since the vector and tensor modes do not arise from the standard cosmology in the linear order.
It is possible to bring information of the phase transition that happened in the early stage of the universe through studying the vector and tensor modes induced from the cosmological defect.

%Weak lensing
We focus on the weak lensing from the vector and tensor modes induced by the non-topological texture.
Photons emitted from the CMB last scattering surface and galaxies are deflected by the foreground scalar, vector, and tensor perturbations, called the CMB lensing and the cosmic shear, respectively \cite{Bartelmann:1999yn,Lewis:2006fu}.
We can decompose these deflection patterns into the parity-even and parity-odd signatures.
The parity-even signal emerged from the scalar, vector, and tensor modes.
On the other hand, the parity-odd mode is induced only from the vector and tensor modes \cite{Stebbins:1996wx,Hirata:2003ka,Yamauchi:2013fra}.
Therefore, the parity-odd mode of the CMB lensing and the cosmic shear, that is, the curl mode and the cosmic shear B-mode, respectively, are a good probe for the cosmological defects such as the texture.
The parity-even modes of the CMB lensing and the cosmic shear which are induced from the first-order scalar mode have been detected with a high signal-to-noise ratio by e.g., the Planck \cite{Ade:2015zua}, the Canada-France Hawaii Telescope Lensing Survey (CFHTLenS) \cite{Kilbinger:2012qz,Benjamin:2012qp,Simpson:2012ra}, and the Dark Energy Survey (DES) \cite{Abbott:2015swa,Becker:2015ilr}.
In previous studies, many parity-odd models have been studied and discussed, e.g., cosmic (super) strings \cite{Yamauchi:2012bc,Yamauchi:2011cu,Namikawa:2013wda}, primordial gravitational waves \cite{Li:2006si,Dodelson:2003bv}, or the second-order perturbation \cite{Sarkar:2008ii,Saga:2015apa,Saga:2016cvt}.
Although the parity-odd mode has not been detected, the prediction of the parity-odd mode for possible sources must become one of the important observable in the future high sensitivity observations.

%In this paper, ...
In this paper, we study the parity-odd signals from the non-topological texture governed by the NLSM with large-$N$ limit.
The outline of this paper is as follows.
In Section \ref{sec: NLSM}, we review and summarize the NLSM with large-$N$ limit.
The NLSM has $N$-component real scalar fields and the non-linearity of these scalar fields induces the vector and tensor modes.
The vector and tensor modes from the NLSM with large-$N$ limit can be determined by solving Einstein equation.
In addition, we give an analytical estimation of the vector and tensor modes.
In Section \ref{sec: WL}, we present the formulation of weak lensing signals.
As mentioned above, we focus on the parity-odd signatures, that is, the curl mode for the CMB lensing and the B-mode for the cosmic shear.
In Section \ref{sec: RAD}, we provide results and discussions.
We also give analytical estimates of the lensing signal and discussions of the detectability of the non-topological texture.
In Section \ref{sec: SUM}, we provide our conclusion.

%============== Section ==============%
\section{Non-linear sigma model}\label{sec: NLSM}
In this section, we review the non-linear sigma model (NLSM), which has the vector and tensor modes originated from the anisotropic stress of scalar fields.
The NLSM can accurately describe cosmological defects with the global $O(N)$ symmetry in the case of $N>2$ \cite{PhysRevLett.63.2625,PhysRevLett.66.3093}.
Throughout this paper, we assume the background metric is given by the Friedman-Robertson-Walker metric as
\begin{eqnarray}
ds^{2} = a(\eta)^{2} \left[ -d\eta^{2} + d\bm{x}^{2} \right] ~,
\end{eqnarray}
where $\eta$ and $a(\eta)$ are the conformal time and the scale factor, respectively.

We focus on the dynamics of real $N$-scalar fields with the Lagrangian which satisfies the global $O(N)$ symmetry:
\begin{eqnarray}
\mathcal{L} =
-\frac{1}{2}\left( \nabla_{\mu}\bm{\Phi}^{t} \right)\left( \nabla^{\mu}\bm{\Phi} \right)
- \frac{\lambda}{4}\left( \bm{\Phi}^{t}\bm{\Phi} - v^{2} \right)^{2}
+ \mathcal{L}_{\rm T} ~, \label{eq: Phi O(N)}
\end{eqnarray}
where we define the array of real $N$-scalar fields as $\bm{\Phi} = \left( \phi_{1}, \phi_{2}, \cdots, \phi_{N} \right)$.
Moreover, $v$ and $\lambda$ are the vacuum expectation value (VEV) and the dimensionless self-coupling parameter, respectively.
The interaction with the thermal environment having the temperature $T$ is represented as $\mathcal{L}_{\rm T} \sim T^{2} \bm{\Phi}^{t}\bm{\Phi}$.
In the case of low temperature, $T \ll v$, the global $O(N)$ symmetry breaks spontaneously to $O(N-1)$ symmetry with the condition $\bm{\Phi}^{t}\bm{\Phi} = v^{2}$.
According to this constraint, the equation of motion for scalar fields is determined from Eq.~(\ref{eq: Phi O(N)}) as
\begin{eqnarray}
\nabla^{\mu}\nabla_{\mu}\beta_{a} + \sum_{b = 1}^{N-1}\left( \nabla^{\mu} \beta_{b}\right)\left( \nabla_{\mu} \beta^{b}\right) \beta_{a} = 0 ~, \label{eq: beta}
\end{eqnarray}
where $\beta_{a}$ is scalar fields normalized by the VEV, namely, $\beta_{a} \equiv \Phi_{a}/v$.
The normalized scalar fields obey the condition $\sum_{a=1}^{N}\beta_{a}\beta^{a} = 1$.
The above equation (\ref{eq: beta}) is called the non-linear sigma model.

By taking the large-$N$ limit in Eq.~(\ref{eq: beta}), the solution of Eq.~(\ref{eq: beta}) in the Fourier space is given as \cite{Horiguchi:2015xsa}
\begin{eqnarray}
\beta_{a}(\bm{k},\eta) =
\sqrt{A_{\nu}}\left( \frac{\eta}{\eta_{\rm ini}}\right)^{3/2} \frac{J_{\nu}(k\eta)}{\left( k\eta \right)^{\nu}} \beta_{a}(\bm{k},\eta_{\rm ini}) ~,
\end{eqnarray}
where $\nu \equiv d\ln{a}/d\ln{\eta} + 1$ and $A_{\nu}\equiv 4\Gamma(2\nu - 1/2)\Gamma(\nu - 1/2)/\left( 3\Gamma(\nu - 1)\right)$.
We assume that $\beta_{a}(\bm{k},\eta_{\rm ini})$ are random gaussian variables.
During the radiation- and matter-dominated eras, the parameter $\nu$ takes $\nu_{\rm rad} = 2$ and $\nu_{\rm mat} = 3$, respectively.
Although the solution of scalar fields $\beta_{a}$ depends on the phase transition time $\eta_{\rm ini}$, the power spectrum of scalar fields is independent of this time \cite{Horiguchi:2015xsa}.
The dimensionless power spectrum for normalized scalar fields can be given as
\begin{eqnarray}
\Braket{\beta_{a}(\bm{k},\eta)\beta^{*}_{b}(\bm{k'},\eta)}
&=& \frac{2\pi^{2}}{k^{3}}\mathcal{P}_{\beta}(k,\eta)
\delta_{ab} (2\pi)^{3}\delta^{3}_{d}(\bm{k}-\bm{k'}) ~, \\
\mathcal{P}_{\beta}(k,\eta) &=& \frac{3 A_{\nu}}{N} (k\eta)^{3} \left( \frac{J_{\nu}(k\eta)}{\left( k\eta \right)^{\nu}} \right)^{2} ~, \label{eq: power beta}
\end{eqnarray}
where the initial power spectrum is determined as (see e.g., Ref.~\cite{Fenu:2009qf})
\begin{eqnarray}
\Braket{\beta_{a}(\bm{k},\eta_{\rm ini})\beta^{*}_{b}(\bm{k'},\eta_{\rm ini})}
= \begin{cases}
	\frac{6\pi^{2}\eta^{3}_{\rm ini}}{N} \delta_{ab} (2\pi)^{3}\delta^{3}_{d}(\bm{k}-\bm{k'}) & (k\eta_{\rm ini} \ll 1) \\
	0 & (k\eta_{\rm ini} \gtrsim1) ~.
  \end{cases}
\end{eqnarray}
The amplitude of the solution is determined to satisfy the condition $\sum_{a = 1}^{N}\beta_{a} \beta^{a} = 1$.
Note that, the configuration of scalar fields is not correlated on sub-horizon scales, i.e., $k\eta_{\rm ini}\gtrsim 1$.
In other words, as expressed in the above equation, the correlation of scalar fields vanishes in these scales.
From Eq.~(\ref{eq: power beta}), we can see that the power spectrum of scalar fields does not depend on the initial time.
Therefore, we have omitted the initial time $\eta_{\rm ini}$ from the argument of the power spectrum.
The energy momentum tensor for scalar fields is written as
\begin{eqnarray}
T^{\phi}_{\mu\nu} = v^{2}\sum_{a}\left[ \left( \partial_{\mu}\beta_{a} \right) \left( \partial_{\nu}\beta^{a} \right) - \frac{1}{2}g_{\mu\nu} \left( \partial_{\lambda}\beta_{a} \right)\left( \partial^{\lambda}\beta^{a} \right)\right] ~.
\end{eqnarray}
The anisotropic stress of scalar fields corresponds to the $(i,j)$ component of the energy momentum tensor.

From here, we derive evolution equations for the vector and tensor metric perturbations with the anisotropic stress of self-ordering scalar fields.
In our study, we work in the Poisson gauge given by
\begin{eqnarray}
ds^{2} = a^{2}(\eta) \left[ -d\eta^{2} + 2\sigma_{i}d\eta dx^{i} + \left( \delta_{ij} + h_{ij}\right)dx^{i}dx^{j}\right] ~,
\end{eqnarray}
where we drop the scalar metric perturbation since we are interested in the vector $\sigma_{i}$ and tensor $h_{ij}$ perturbations.
Due to the gauge conditions, the vector and tensor perturbations satisfy $\sigma^{i}{}_{,i} = h^{ij}{}_{,i} = 0$.

The Einstein equations for the vector $\sigma_{\rm V}$ and tensor $h_{\rm T}$ perturbations in the Fourier space are given as
\begin{eqnarray}
k\left[ \dot{\sigma}_{\rm V}(\bm{k},\eta) + 2\mathcal{H}\sigma_{\rm V}(\bm{k},\eta) \right]
&=& \frac{8\pi}{m_{\rm pl}^{2}} \pi^{\phi}_{\rm V}(\bm{k},\eta) ~, \\
\ddot{h}_{\rm T}(\bm{k},\eta) + 2\mathcal{H}\dot{h}_{\rm T}(\bm{k},\eta) +k^{2}h_{\rm T}(\bm{k},\eta)
&=& \frac{8\pi}{m_{\rm pl}^{2}} \pi^{\phi}_{\rm T}(\bm{k},\eta) ~, \label{eq: tensor}
\end{eqnarray}
where a dot denotes the derivative with respect to the conformal time.
Anisotropic stresses for the vector and tensor modes can be given by the product of scalar fields as
\begin{eqnarray}
\pi^{\phi}_{\rm V}(\bm{k},\eta)&=& \int{\frac{d^{3}\bm{q}}{(2\pi)^{3}}}\int{\frac{d^{3}\bm{p}}{(2\pi)^{3}}}\delta^{3}_{d}(\bm{k} - \bm{q} - \bm{p})
\left[ \frac{v^{2}}{2} \sqrt{1-\mu^{2}} q (k - 2q \mu) \right]
\sum_{a}\beta_{a}(\bm{q},\eta) \beta^{a}(\bm{p},\eta) ~, \label{eq: NLSM vec}\\
\pi^{\phi}_{\rm T}(\bm{k},\eta) &=& \int{\frac{d^{3}\bm{q}}{(2\pi)^{3}}}\int{\frac{d^{3}\bm{p}}{(2\pi)^{3}}}\delta^{3}_{d}(\bm{k} - \bm{q} - \bm{p})
\left[ v^{2} \left( 1- \mu^{2} \right) q^{2} \right]
\sum_{a}\beta_{a}(\bm{q},\eta) \beta^{a}(\bm{p},\eta) ~, \label{eq: NLSM ten}
\end{eqnarray}
where we define $\mu \equiv \bm{\hat{k}}\cdot\bm{\hat{q}}$.
In order to predict the weak lensing signal, we define the dimensionless unequal-time power spectra for the vector and tensor modes which are defined as
\begin{eqnarray}
\Braket{\xi_{\rm X}(\bm{k},\eta)\xi^{*}_{\rm X}(\bm{k'},\eta')} = (2\pi)^{3}\delta^{3}_{d}(\bm{k}-\bm{k'}) \frac{2\pi^{2}}{k^{3}} \mathcal{P}_{\rm X}(k,\eta,\eta') ~,
\end{eqnarray}
where $\xi_{\rm X}$ denotes the vector ($\xi_{\rm X} = \sigma_{\rm V}$) and tensor ($\xi_{\rm X} = h_{\rm T}$) modes.
We can solve evolution equations for the vector and tensor modes in Eqs.~(\ref{eq: NLSM vec}) and (\ref{eq: NLSM ten}) straightforwardly.
By using solutions of the vector and tensor modes, we can write down the dimensionless unequal-time power spectrum during the matter-dominated era ($\nu = 3$) as
\begin{eqnarray}
\mathcal{P}_{\rm X}(k\eta, k\eta') & = &
\mathcal{A}
\int^{\infty}_{-\infty} d\ln{q_{k}}
\int^{1}_{-1} d\mu
\mathcal{F}_{\rm X}(q_{k}, \mu , k\eta)
\mathcal{F}_{\rm X}(q_{k}, \mu , k\eta') ~,\\
\mathcal{A} & = & 144\pi^{2} A^{2}_{3} \epsilon^{2}_{v} ~, \notag \\
&\approx& 1.22\times 10^{7} \epsilon^{2}_{v} ~, \\
\mathcal{F}_{\rm V}(q_{k},\mu, x) & = &
\sqrt{1-\mu^2} \left(1-2q_{k}\mu\right) q_{k}^{5/2}\frac{1}{x^{4}}
\int^{x}_{0} dx_{1}\ 
x_{1}^{7}
\frac{J_{3}(q_{k} x_{1})}{(q_{k} x_{1})^{3}}
\frac{J_{3}(p_{k} x_{1})}{(p_{k} x_{1})^{3}} ~, \label{eq: Fv}\\
\mathcal{F}_{\rm T}(q_{k},\mu, x) & = &
2(1-\mu^2) q_k^{7/2}\frac{1}{x^{3}}
\int^{x}_0 dx_{1}\ \left[ x x_{1}G(x, x_{1}) \right]x_{1}^{4}\frac{J_3(q_kx_{1})}{(q_kx_{1})^3}\frac{J_3(p_kx_{1})}{(p_kx_{1})^3} ~,
\end{eqnarray}
where $q_{k}\equiv q/k$, $p_{k} \equiv p/k$, $x\equiv k\eta$, and $G(x,x_{1}) = xx_{1}\left( j_{1}(x_{1})n_{1}(x) - j_{1}(x)n_{1}(x_{1})\right)$ is the Green function for the evolution equation of the tensor mode (\ref{eq: tensor}), and $J_{\nu}(x)$, $j_{\nu}(x)$, and $n_{\nu}(x)$ are the Bessel function, the spherical Bessel function, and the spherical Neumann function, respectively.
The shape of the unequal-time power spectrum does not depend on the theoretical parameters such as $N$ and $v$.
These parameters change only the amplitude of the power spectrum and appear through a special combination of $N^{-1}v^{4}$.
Therefore, in this paper, we define a new parameter through the combination of theoretical parameters as
\begin{eqnarray}
\epsilon^{2}_{v} \equiv N^{-1}\left( v/ m_{\rm pl}\right)^{4} ~.
\end{eqnarray}
In this paper, for simplicity, we evaluate the weak lensing signal by using the power spectrum during the matter-dominated era.
The correction to the radiation component should be small since lensing signals are mainly contributed from the perturbations at late times of cosmic evolution.

\begin{figure}[t]
\begin{center}
\includegraphics[width=0.6\textwidth]{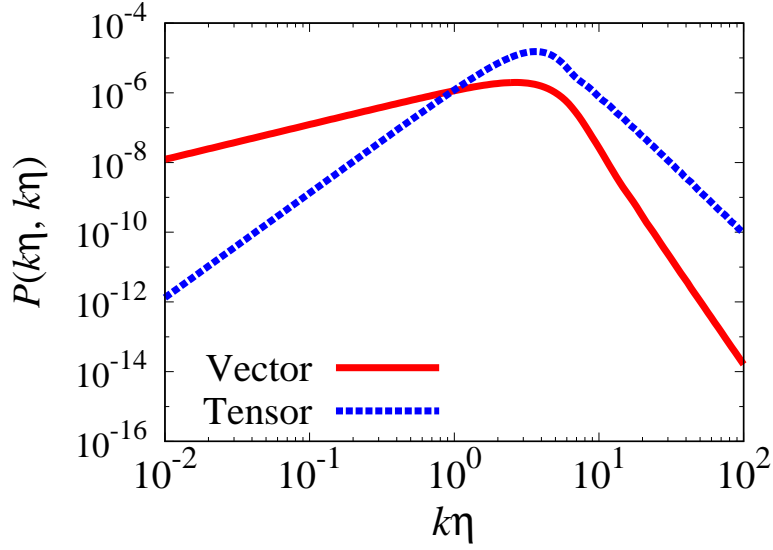}
\end{center}
\caption{%
Power spectra $\mathcal{P}(k\eta, k\eta)$ for the vector and tensor modes from the NLSM.
For the presentation purpose, we set $\mathcal{A} = 1$ in this figure.
Due to the convolution of scalar fields in Eqs.~(\ref{eq: NLSM vec}) and (\ref{eq: NLSM ten}), the peak moves to smaller scales than the horizon scale where $k\eta = 1$.
}
\label{fig: auto power}
\end{figure}

We depict the dimensionless equal-time power spectrum for the vector and tensor modes in Fig.~\ref{fig: auto power}.
%===
%===
We can see that on super (sub) horizon scales, the vector (tensor) mode is greater than the tensor (vector) mode.
In the following section, in order to discuss the angular power spectra of the curl and B modes, we evaluate the asymptotic power on small scales.
From here, we estimate the asymptotic power of the dimensionless equal-time power spectrum on sub-horizon scales as follows.
At first, let us see the vector mode.
By integrating Eq.~(\ref{eq: Fv}), we obtain the notation of $\mathcal{F}_{\rm V}(q_{k},\mu, x)$ exactly as
\begin{equation}
\mathcal{F}_{\rm V}(q_{k},\mu, x)=\sqrt{1-\mu^2}q_k^{-1/2}p_k^{-3}x^{-3}\left(q_kJ_2(q_kx)J_3(p_kx)-p_kJ_3(q_kx)J_2(p_kx)\right) ~. \label{eq:Fvi}
\end{equation}
Using the approximations for the Bessel function, $J_\nu(x\ll \nu)\propto x^{\nu}$ and $J_\nu(x\gg\nu)\propto x^{-1/2}{\rm cos}x$, and assuming a cutoff scale $1/x$, we can integrate the auto-power spectrum for the vector mode as 
\begin{eqnarray}
\mathcal{P}_{\rm V}(x,x)&\propto& x^{-6}\int^{1/x} dq_k
\Bigl[ p_{k}^{-6}J_{3}^{2}(p_{k}x)J_{2}^{2}(q_{k}x) \notag \\
&& -2p_{k}^{-5}q_{k}^{-1}J_{2}(p_{k}x)J_{3}(p_{k}x)J_{2}(q_{k}x)J_{3}(q_{k}x)+p_{k}^{-4}q_{k}^{-2}J_{2}^{2}(p_{k}x)J_{3}^{2}(q_{k}x) \Bigr] \notag\\
&\sim&\alpha_{1}x^{-8}+\alpha_{2}x^{-7}+\alpha_{3}x^{-6} \notag\\
&\propto&x^{-6} ~,
\end{eqnarray}
where $\alpha_{1},\ \alpha_{2}$ and $\alpha_{3}$ are constants.
Therefore the $k\eta$ dependence of $\mathcal{P}_{\rm V}(k\eta,k\eta)$ is $\propto (k\eta)^{-6}$.
Next, we see the tensor mode.
Here we find the most dominant term of $\mathcal{F}_{\rm T}(q_k,\mu,x)$, that is, the highest power of $x$ and $x_{1}$, by considering the integrand as
\begin{eqnarray}
\mathcal{F}_{\rm T}(q_k,\mu,x)
&\sim&p_{k}^{-3}q_{k}^{-1/2}x^{-3}\int^{x }dx_{1}\left[(x_{1}-x){\rm cos}(x-x_{1})+(1+xx_{1}){\rm sin}(x-x_{1})\right]x_{1}^{-2}J_{3}(q_{k}x_{1})J_{3}(p_{k}x_{1})\notag\\
&\sim&p_{k}^{-3}q_{k}^{-1/2}x^{-2}\int^{x\rightarrow1/q_{k}} dx_{1}x_{1}^{-1}{\rm sin}(x-x_{1})J_{3}(q_{k}x_{1})J_{3}(p_{k}x_{1})\notag\\
&\sim&p_{k}^{-3}q_{k}^{-1/2}x^{-2}q_{k}J_{3}(p_{k}/q_{k}),
\end{eqnarray}
where we have assumed $p_k>q_k$ and we can obtain the expression for the case $p_k<q_k$ in the same way. Now we are able to calculate the $k\eta$ dependence of $\mathcal{P}_{\rm T}(x,x)$ as
\begin{eqnarray}
\mathcal{P}_{\rm T}(x,x)&\propto&\int dq_{k}q_{k}^{-1}\left[p_{k}^{-3}q_{k}^{-1/2}x^{-2}q_{k}J_{3}(p_{k}/q_{k})\right]^{2} \notag\\
&\propto&x^{-4}\propto (k\eta)^{-4} ~.
\end{eqnarray}
Here we have obtained the $k\eta$ dependence of the dimensionless equal-time power spectrum for vector and tensor modes as $\mathcal{P}_{\rm V}(k\eta,k\eta)\propto (k\eta)^{-6}$ and $\mathcal{P}_{\rm T}(k\eta,k\eta)\propto (k\eta)^{-4}$, respectively. These spectra leave various trails on physical values and these estimations enable us to predict their analytic forms. 
%============== Section ==============%
\section{Weak lensing}\label{sec: WL}
In this section, we give a review about the relation between weak lensing signals and vector and tensor perturbations from the textures following Refs.~\cite{Yamauchi:2012bc,Yamauchi:2013fra}.
We pull parity-odd signals from the CMB lensing and the cosmic shear which are called the curl and B modes, respectively.
In the following subsection, we present details about the curl and B modes.
%--------------------- Subsection ---------------------%
\subsection{CMB lensing curl mode}
CMB photons are deflected by foreground scalar, vector, and tensor perturbations.
We decompose the deflection angle of CMB photons projected on the celestial sphere $\Delta_{a}(\bm{\hat{n}})$ into the gradient ($\phi(\bm{\hat{n}})$) and curl ($\varpi(\bm{\hat{n}})$) modes as
\begin{eqnarray}
\Delta_{a}(\bm{\hat{n}}) = \nabla_{a}\phi (\bm{\hat{n}}) + \left( \nabla_{b}\varpi (\bm{\hat{n}}) \right)\epsilon^{b}{}_{a} ~,
\end{eqnarray}
where $\epsilon^{b}{}_{a}$ is the covariant dimensional Levi-Civita tensor.
Note that latin characters started from $a$, $b$, $\cdots$ in the above relation denote the azimuthal and polar angles denoted as $\theta$ and $\phi$, respectively.
From here, we drop the gradient mode since, in this paper, we are interested in the curl mode.

In order to relate the curl mode and the angular power spectrum, we solve the geodesic equation in the perturbed spacetime.
By solving the perturbed geodesic equation, the curl mode can be expressed by using the metric perturbations of the vector and tensor modes as
\begin{eqnarray}
\varpi^{:a}{}_{:a} = - \int^{\chi_{\rm S}}_{0}d\chi \frac{\chi_{\rm S} - \chi}{\chi\chi_{\rm S}} \left[ \frac{d}{d\chi} \left( \chi \Omega^{a}{}_{:b} \epsilon^{b}{}_{a}\right)\right] ~, \label{eq: varpi omega}
\end{eqnarray}
where $\chi$ is the comoving distance measured from the observer at the origin and $\chi_{\rm S}$ is the comoving distance at the sources.
$\Omega^{a}$ in Eq.~(\ref{eq: varpi omega}) includes the vector and tensor perturbations as
\begin{eqnarray}
\Omega_{a} = \left( -\sigma_{i} + h_{ij}e^{j}_{\chi}\right)e^{i}_{a}~,
\end{eqnarray}
where $e^{i}_{\chi}$ and $e^{i}_{a}$ are the orthogonal spacelike basis along the light ray.
We expand the curl mode by using the spherical harmonics and define the angular power spectrum for the curl mode as
\begin{eqnarray}
\varpi(\bm{\hat{n}}) &=& \sum_{\ell ,m} \varpi_{\ell ,m}Y_{\ell,m}(\bm{\hat{n}}) ~, \\
C^{\varpi\varpi}_{\ell} &=& \frac{1}{2\ell + 1}\sum_{m=-\ell}^{\ell} \Braket{\varpi_{\ell ,m}\varpi^{*}_{\ell,m}} ~.
\end{eqnarray}

Finally, we obtain the angular power spectrum of the curl mode in terms of the vector (${\rm X} = {\rm V}$) and tensor (${\rm X} = {\rm T}$) perturbations as
\begin{eqnarray}
C^{({\rm X})\varpi\varpi}_{\ell} = 
4\pi \int^{\infty}_{0} \frac{dk}{k}
\int^{\chi_{S}}_{0}{k d\chi}\int^{\chi_{S}}_{0}{k d\chi'}
\mathcal{S}^{({\rm X})}_{\varpi, \ell}(k\chi)\mathcal{S}^{({\rm X})}_{\varpi, \ell}(k\chi') \mathcal{P}_{\rm X}(k, \eta_{0}-\chi,\eta_{0}-\chi') ~, \label{eq: cl curl}
\end{eqnarray}
where $\mathcal{P}_{\rm X}(k, \eta, \eta')$ denotes the dimensionless unequal-time power spectrum of metric perturbations.
$\mathcal{S}^{({\rm X})}_{\varpi, \ell}(k\chi)$ is the weight function defined as
\begin{eqnarray}
\mathcal{S}^{({\rm V})}_{\varpi, \ell}(x) &=& \sqrt{\frac{(\ell-1)!}{(\ell+1)!}} \frac{j_{\ell}(x)}{x} ~, \\
\mathcal{S}^{({\rm T})}_{\varpi, \ell}(x) &=& \frac{1}{2}\frac{(\ell-1)!}{(\ell+1)!}\sqrt{\frac{(\ell + 2)!}{(\ell -2)!}} \frac{j_{\ell}(x)}{x^{2}} ~.
\end{eqnarray}
In the case of the CMB lensing, the comoving distance to the source $\chi_{\rm S}$ corresponds to that to the CMB last scattering surface.

We assume that the curl-mode lensing potential is reconstructed by using the quadratic estimator \cite{Cooray:2005hm,Namikawa:2011cs}.
In this case, the CMB lensing noise arises from the lensing reconstruction noise from the cosmic variance of the lensed CMB fluctuations.
We assume an ideal experiment for the CMB lensing throughout this paper and neglect instrumental noise.
Consequently, the noise of the CMB lensing is limited by the reconstruction noise due to the quadratic estimator.

%--------------------- Subsection ---------------------%
\subsection{Cosmic shear B-mode}
The intrinsic shape of galaxies is deformed by foreground perturbations.
The deformation pattern is characterized by the reduced shear \cite{Seitz:1994xf,Lewis:2006fu}.
The geodesic deviation equation describes the deformation of the shape of galaxies.
By solving the geodesic deviation equation, we can relate the reduced shear and the vector and tensor perturbations as \cite{Yamauchi:2012bc,Yamauchi:2013fra}
\begin{eqnarray}
g = -\frac{1}{2}\int^{\chi_{\rm S}}_{0}{d}\chi \frac{\chi_{\rm S} - \chi}{\chi\chi_{\rm S}} \left[ \nabla_{a}\nabla_{b}\Upsilon - \frac{d}{{d}\chi} \left( \chi \nabla_{b}\Omega_{a}\right)\right] e^{a}_{+}e^{b}_{+} - \frac{1}{4}\left[ h_{ab}e^{a}_{+}e^{b}_{+}\right]^{\chi_{\rm S}}_{0} ~,
\end{eqnarray}
where $\Upsilon$ contains the scalar, vector, and tensor modes as
\begin{eqnarray}
\Upsilon = -\left( \Psi + \Phi\right) -\sigma_{i}e^{i}_{\chi} + \frac{1}{2}h_{ij}e^{i}_{\chi}e^{j}_{\chi} ~.
\end{eqnarray}
Note that $\Upsilon$ does not appear in the cosmic shear B-mode but in the cosmic shear E-mode.
Therefore, we do not focus on $\Upsilon$ when we study the cosmic shear B-mode.
Because the reduced shear is a spin-2 variable, we can expand the reduced shear according to the spin-2 spherical harmonics as 
\begin{eqnarray}
g(\bm{\hat{n}}) = \sum_{\ell, m} \left( E_{\ell m}+ iB_{\ell m}\right) {}_{+2}Y_{\ell m}(\bm{\hat{n}}) ~,
\end{eqnarray}
where we split multipole coefficients into E and B modes by using the parity.
Hereafter, we focus on the cosmic shear B-mode and drop the E-mode.
As well as the CMB lensing, the angular power spectrum of the B mode is defined as
\begin{eqnarray}
C^{BB}_{\ell} = \frac{1}{2\ell + 1} \sum^{\ell}_{m = -\ell}\Braket{B_{\ell m}B^{*}_{\ell m}} ~.
\end{eqnarray}
By solving the perturbed geodesic deviation equation, we can relate the angular power spectrum of the B mode and the vector or tensor metric perturbations as
\begin{eqnarray}
C^{({\rm X})BB}_{\ell} = \left[ \frac{1}{4} \frac{(\ell + 2)!}{(\ell - 2)!}\right]4\pi \int^{\infty}_{0}\frac{dk}{k}
\int^{\infty}_{0}{k d\chi}\int^{\infty}_{0}{k d\chi'}
\mathcal{S}^{({\rm X})}_{B,\ell}(k,\chi)\mathcal{S}^{({\rm X})}_{B,\ell}(k,\chi') \mathcal{P}_{\rm X}(k, \eta_{0}-\chi,\eta_{0}-\chi') ~, \label{eq: cl BB}
\end{eqnarray}
where weight functions are defined as
\begin{eqnarray}
\mathcal{S}^{({\rm V})}_{B,\ell}(k, \chi) &=& \sqrt{\frac{(\ell -1)!}{(\ell +1)!}}
\int^{\infty}_{\chi}d\chi_{S}\frac{N(\chi_{S})}{N_{\rm g}}\frac{j_{\ell}(k\chi)}{k\chi} ~,\\
\mathcal{S}^{({\rm T})}_{B,\ell}(k, \chi) &=& \frac{1}{2}\frac{(\ell -1)!}{(\ell +1)!}\sqrt{\frac{(\ell+2)!}{(\ell-2)!}}
\left[
	\int^{\infty}_{\chi}d\chi_{S}\frac{N(\chi_{S})}{N_{\rm g}} \frac{j_{\ell}(k\chi)}{(k\chi)^{2}}
\right] \notag \\
&&+ \frac{1}{2}\sqrt{\frac{(\ell - 2)!}{(\ell + 2)!}}\frac{N(\chi)}{N_{\rm g}}\left( j'_{\ell}(k\chi) + 2\frac{j_{\ell}(k\chi)}{k\chi}\right) ~.
\end{eqnarray}
To investigate the cosmic shear signals, we need the distribution of galaxies $N(\chi)$, which should be determined by observations.
Here we assume the following form:
\begin{eqnarray}
N(\chi){d}\chi = N_{\rm g}\frac{3}{2}\frac{z^{2}}{(0.64 z_{\rm m})^{3}}\exp\left[ -\left( \frac{z}{0.64z_{\rm m}}\right)^{3/2}\right]{d}z ~,
\end{eqnarray}
where $z_{\rm m}$ is the mean redshift, and $N_{\rm g}$ is the number of galaxies per square arc-minute.
In our study, we assume three ongoing and forthcoming survey designs, that is, the Subaru Hyper-Suprime Cam (HSC) \cite{HSC:coll}, the Square Kilometer Array (SKA) \cite{Brown:2015ucq}, the Large Synoptic Survey Telescope (LSST) \cite{2009arXiv0912.0201L}.
Individual experimental specifications are summarized in Table.~\ref{tab: designs}.
\begin{table}[t]
\begin{center}
{\tabcolsep = 3mm
\begin{tabular}{| c || c | c | c |} \hline
 & $f_{\rm sky}$ & $z_{\rm m}$ & $N_{\rm g} {\rm [arcmin^{-2}]}$ \\ \hline
HSC & $0.05$ & $1.0$ & $35$ \\ \hline
SKA & $0.75$ & $1.6$ & $10$ \\ \hline
LSST & $0.5$ & $1.5$ & $100$ \\ \hline
\end{tabular}}
\end{center}
\caption{The experimental specifications of the HSC, SKA, and LSST.}
\label{tab: designs}
\end{table}
We assume that the noise of the cosmic shear is the shot noise originated from the intrinsic shape of galaxies written as
\begin{eqnarray}
N^{BB}_{\ell} = \frac{\Braket{\gamma^{2}_{\rm int}}}{3600 N_{\rm g}\left( 180/\pi \right)^{2}} ~, \label{eq: B mode noise}
\end{eqnarray}
where $\Braket{\gamma^{2}_{\rm int}}^{1/2}$ is the root-mean square ellipticity of galaxies, which is determined about $0.3$ in Ref.~\cite{Bernstein:2001nz}.

Note that without the dependence of the distribution of galaxies, i.e., $N(\chi) = {\rm const.}$, there is the relation between the CMB lensing curl-mode and the cosmic shear B-mode power spectra as \cite{Yamauchi:2013fra}
\begin{eqnarray}
C^{\varpi\varpi}_{\ell} = 4\frac{(\ell-2)!}{(\ell+2)!} C^{BB}_{\ell} ~. \label{eq: cl curl and B}
\end{eqnarray}
We use this relation in the following section to obtain the asymptotic scaling of the angular power spectra.
 
Before closing this section, we mention our treatment about the unequal-time power spectrum.
To calculate weak lensing signals, we need to use the unequal-time power spectrum for the vector and tensor modes.
For simplicity, to perform the multiple integration, we assume the case of the totally coherent model \cite{Magueijo:1996px,Durrer:1997rh,Durrer:2001cg,Yamauchi:2012bc} throughout this paper.
In other words, we can write the unequal-time power spectrum as
$\mathcal{P}_{\rm X}(k\eta, k\eta') = \sqrt{\mathcal{P}_{\rm X}(k\eta, k\eta)\mathcal{P}_{\rm X}(k\eta', k\eta')}$.
This assumption makes the computation of the angular power spectrum easy.
From Eqs.~(\ref{eq: cl curl}) and (\ref{eq: cl BB}), the unequal-time power spectrum is multiplied by the weight functions, which correspond to the spherical Bessel functions.
The dominant contributions of the integrands on the angular power spectrum would be $\ell \sim k\eta$ since the spherical Bessel function $j_{\ell}(x)$ rapidly decays at $x > \ell$.
We will show that it is sufficient to assume the totally coherent model on small scales by using the small-angle approximation, i.e., the Limber approximation.
Therefore, the totally coherent model is a good approximation on small scales but not on large scales.
We will discuss the detail of the effect of the totally coherent model in the next section.

%============== Section ==============%
\section{Results and discussions}\label{sec: RAD}
\begin{figure}[t]
\begin{center}
\includegraphics[width=0.49\textwidth]{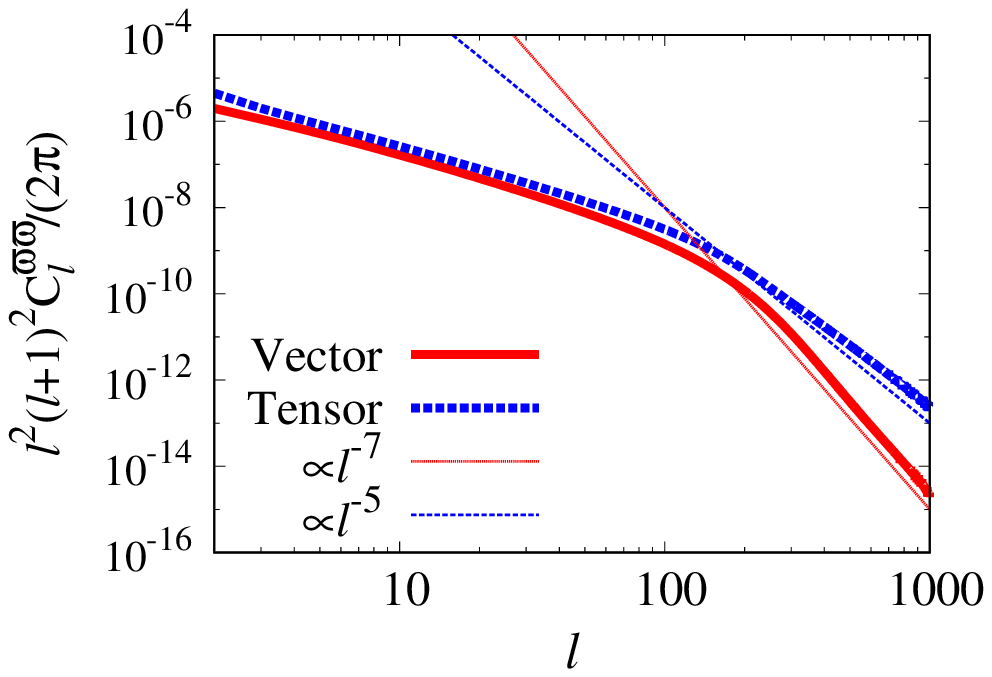}
\includegraphics[width=0.49\textwidth]{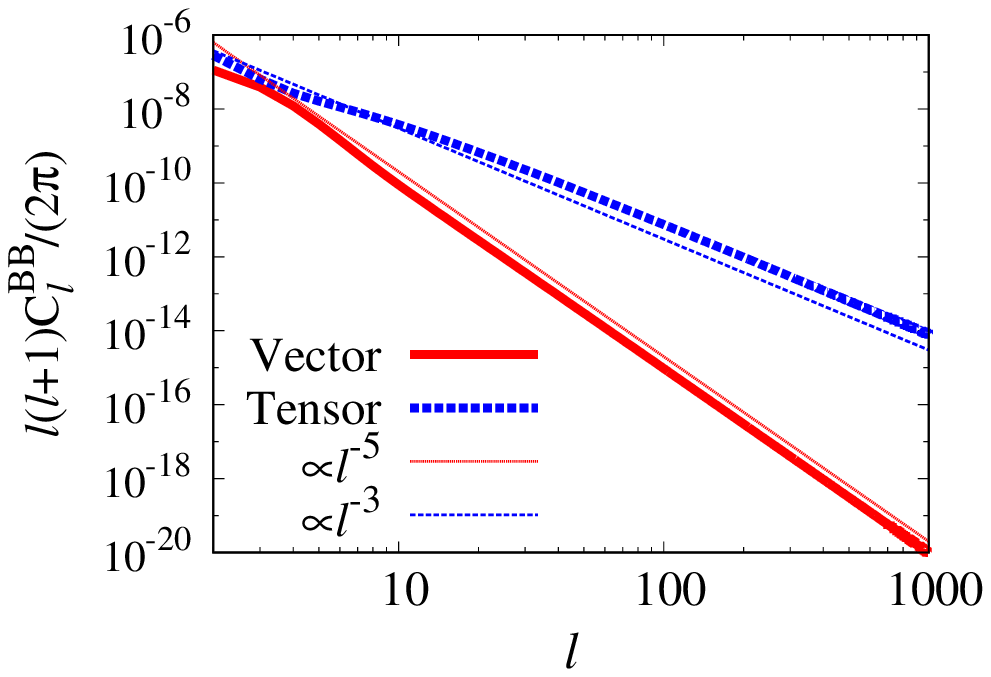}
\end{center}
\caption{%
Left: The angular power spectrum of the CMB lensing curl-mode from vector- and tensor-modes of the texture.
Right: The angular power spectrum of the cosmic shear B-mode from vector- and tensor-modes of the texture by assuming the observation as LSST.
For the same reason in Fig.~\ref{fig: auto power}, we set the theoretical parameter set $\mathcal{A} = 1$ in both figures.
}
\label{fig: lensing 1}
\end{figure}

In this section, we present our main results and give discussions.
In Fig.~\ref{fig: lensing 1}, we show weak lensing signals from the global texture modeled by the NLSM.
%===
%===
We can find that the contribution to the lensing signal is dominated by the tensor mode.
This is because the spherical Bessel function in Eqs.~(\ref{eq: cl curl}) and (\ref{eq: cl BB}) projects on the angular power spectrum around $\ell \sim k\eta$ which corresponds to sub-horizon scales.
In Fig.~\ref{fig: auto power}, the tensor mode has larger amplitude than the vector mode on sub-horizon scales.
Therefore, the angular power spectra of the curl and B modes are dominated by the tensor mode.
Moreover, the difference between the vector and tensor contributions on the lensing signal is greater at low redshift observation.
Note that the CMB B-mode polarization from the tensor mode of the texture has almost the same amplitude \cite{Fenu:2013tea}.

The CMB lensing curl-mode from the texture has a broken power at $\ell \approx 200$ which is smaller scale compared with the standard peak of the scalar lensing potential or the lensing from the primordial gravitational waves \cite{Yamauchi:2013fra,Saga:2015apa}.
This is because the peak of the power spectrum from the NLSM does not correspond to the horizon scale but slightly smaller scale due to the nonlinearity (see Fig.~\ref{fig: auto power} or Ref.~\cite{Durrer:2014raa}).
On large scales ($\ell \lesssim 200$), the power of the angular power spectra from the vector and tensor modes is proportional to $\ell^{-2}$.

Moreover, we can obtain the analytic power on small scales $(\ell \gg 1)$ by using the small-angle approximation as follows,
\begin{eqnarray}
C^{({\rm X})\varpi\varpi }_{\ell} &\propto& \int^{\infty}_{0} \frac{dk}{k}
\int^{\chi_{S}}_{0}{k d\chi}\int^{\chi_{S}}_{0}{k d\chi'}
\mathcal{S}^{(X)}_{\varpi, \ell}(k\chi)\mathcal{S}^{(X)}_{\varpi, \ell}(k\chi') \mathcal{P}_{X}(k, \eta_{0}-\chi,\eta_{0}-\chi') \notag \\
&\propto& \frac{1}{\ell^{5}}\int^{\chi_{\rm S}}_{0}{d\chi}\frac{1}{\chi} \mathcal{P}_{\rm X}\left( \ell (\eta_{0}-\chi)\chi^{-1},\ell (\eta_{0}-\chi)\chi^{-1}\right) ~, \label{eq: Limber}
\end{eqnarray}
where we assume the large-$\ell$ limit to provide the above relation and we use the so-called Limber approximation.
In the above equation, when the multipole is quite large, the contribution from the power spectrum is mainly coming from the sub-horizon power, that is, $k\eta \gg 1$.
From Sec.~\ref{sec: NLSM}, we find that the power spectrum on large multipoles $(\ell \gg 1)$ for the vector and tensor modes is therefore proportional to $(k\eta)^{-6}$ and $(k\eta)^{-4}$, respectively.
We can derive the asymptotic power of the weak lensing curl mode as $\ell^{4}C^{({\rm V})\varpi\varpi}_{\ell} \propto \ell^{-7}$ and $\ell^{4}C^{({\rm T})\varpi\varpi}_{\ell} \propto \ell^{-5}$.
From Eq.~(\ref{eq: cl curl and B}), angular power spectra of the CMB lensing and cosmic shear are related as $C^{\varpi\varpi}_{\ell} \propto \ell^{-4} C^{BB}_{\ell}$,
the asymptotic power of the B-mode cosmic shear can be given as $\ell^{2}C^{({\rm V})BB}_{\ell} \propto \ell^{-5}$ and $\ell^{2}C^{({\rm T})BB}_{\ell} \propto \ell^{-3}$.
We can see these asymptotic powers from Fig.~\ref{fig: lensing 1}.
Note that observed lensing signal is the sum of the vector and tensor modes, i.e., $C^{(\rm tot)}_{\ell} = C^{({\rm V})}_{\ell} + C^{({\rm T})}_{\ell}$.

From here, we discuss the detectability of the texture by using the weak lensing signals.
In the case of the CMB lensing, we consider the noise spectrum that is due to the cosmic variance of the CMB, so called the CMB reconstruction noise, assuming a noiseless instrument following Ref.~\cite{Namikawa:2011cs}.
The CMB reconstruction noise mainly depends on the number of available multipoles.
Throughout this paper, we use the lensed and unlensed CMB angular power spectrum up to $\ell_{\rm max} = 3000$ when computing the reconstruction noise.
On the other hand, the noise spectrum of the cosmic shear observations is determined by the shot noise given by Eq.~(\ref{eq: B mode noise}).

\begin{figure}[t]
\begin{center}
\includegraphics[width=0.7\textwidth]{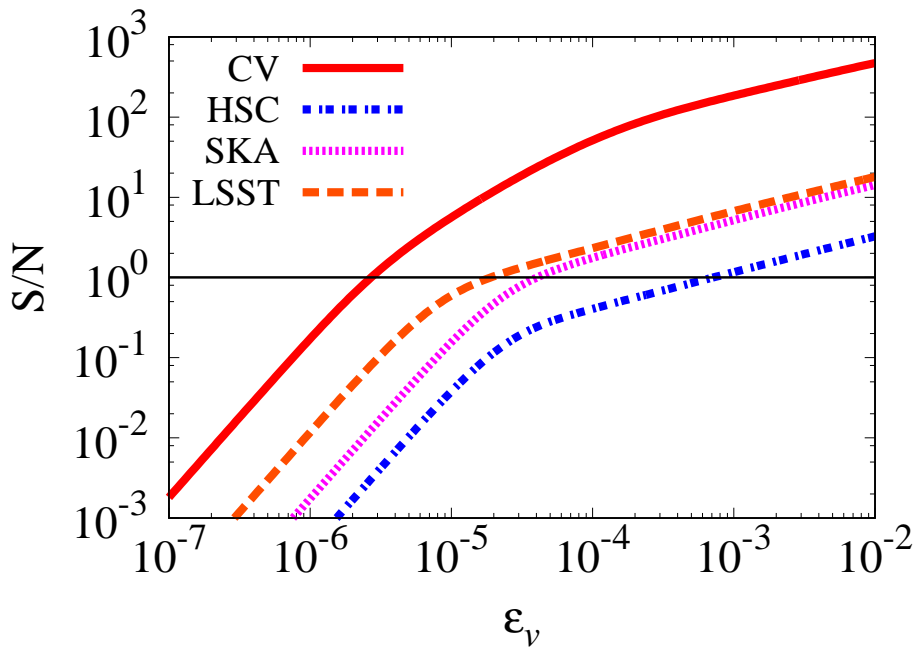}
\end{center}
\caption{%
The signal-to-noise ratio by varying the theoretical parameter $\epsilon_{v}$.
In the case of the CMB lensing denoted as ``CV'' in this figure, the noise spectrum is assumed the CMB reconstruction noise \cite{Namikawa:2011cs}.
In the cosmic shear case denoted as HSC, SKA, and LSST, we assume the shot noise originated from the intrinsic shape of galaxies in Eq.~(\ref{eq: B mode noise}).
In the case of the CMB lensing, we assume the lensing reconstruction noise without the instrumental noise, namely, the cosmic-variance limited noise denoted as CV in this figure.
Moreover, for the cosmic shear experiment, we show the signal-to-noise ratio resulting from the HSC, SKA, and LSST experiments.
We also show the vertical solid line which corresponds to $S/N = 1$.
}
\label{fig: SN v4}
\end{figure}

We estimate the signal-to-noise ratio as
\begin{eqnarray}
\left( \frac{\rm S}{\rm N}\right)_{< \ell} &=& \left[ \sum_{\ell'=2}^{\ell} \left( \frac{C_{\ell'}}{\Delta C_{\ell'}}\right)^{2}\right]^{1/2} ~, \label{eq: SN def1} \\
\Delta C_{\ell} &=& \sqrt{\frac{2}{(2\ell + 1)f_{\rm sky}}} \left( C_{\ell} + N_{\ell}\right) ~. \label{eq: SN def2}
\end{eqnarray}
In Fig.~\ref{fig: SN v4}, we show the relation between the signal-to-noise ratio and the theoretical parameter $\epsilon_{v}$.
%===
%===
We can find that the ultimate experiment of the CMB lensing without including the instrumental noise can set an upper limit on the theoretical parameter related to the VEV as $\epsilon_{v} \sim 2.7\times 10^{-6}$.

\begin{figure}[t]
\begin{center}
\includegraphics[width=0.7\textwidth]{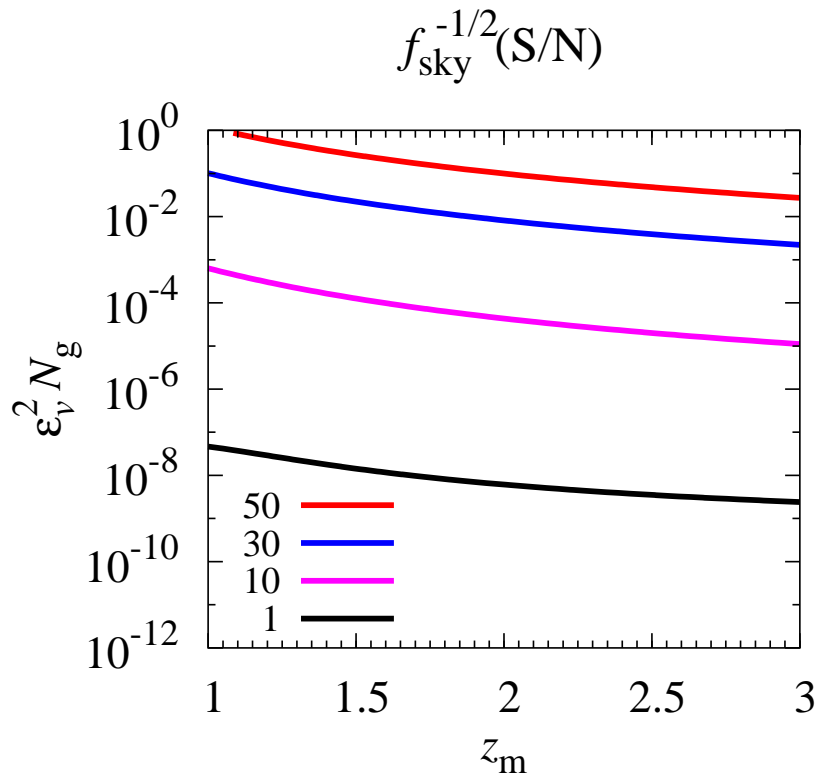}
\end{center}
\caption{%
The signal-to-noise ratio with the factor $f^{-1/2}_{\rm sky}$ as the function of two parameters $\epsilon^{2}_{v}N_{\rm g}$ and $z_{\rm m}$.
This figure shows contours which corresponds to $f^{-1/2}_{\rm sky} (S/N) =$ $1$, $10$, $30$, and $50$.
We set the maximum multipole to estimate the signal-to-noise ratio as $\ell_{\rm max} = 1000$.
}
\label{fig: SN cont}
\end{figure}

Constraints from the cosmic shear are much weaker than those from the CMB lensing.
This is because signals of the cosmic shear are strongly suppressed on small scales.
However, contrary to the CMB lensing observation, the signal-to-noise ratio of cosmic shear experiments depends on parameters of the experimental specification.
Fortunately, the theoretical parameter $\epsilon_{v}$ changes only the amplitude of the angular power spectrum, namely, $C_{\ell} \propto \epsilon^{2}_{v}$.
From the definition of the signal-to-noise ratio (\ref{eq: SN def1}) and (\ref{eq: SN def2}),
the signal-to-noise ratio therefore depends on the special combination $\epsilon^{2}_{v} N_{\rm g}$.
In Fig.~\ref{fig: SN cont}, we show the relation between the signal-to-noise ratio and $\epsilon^{2}_{v} N_{\rm g}$ and the mean redshift.
%===
%===
From this result, we can give a rough estimation of the signal-to-noise ratio as the function of $\epsilon^{2}_{v}N_{\rm g}$ and $z_{\rm m}$, such as $S/N \propto f^{1/2}_{\rm sky} \left[ \epsilon^{2}_{v}N_{\rm g} \right]^{0.2} z^{0.7}_{\rm m}$ for the cosmic shear observation.
According to this estimation, in order to improve the detectability, we need to push $z_{\rm m}$ to higher redshift rather than adding the number of galaxies $N_{\rm g}$ since the signal-to-noise ratio is sensitive to the mean redshift rather than the observing number of galaxies.

Before closing this section, we discuss the validity of the assumption, that is, the totally coherent model.
Under the Limber approximation presented in Eq.~(\ref{eq: Limber}), the power spectrum on small scales is determined by the equal-time power spectrum, which is the same as the totally coherent model.
Therefore, the totally coherent model is valid on small scales.

In Fig.~\ref{fig: lensing 1}, we can see that the Limber approximation can explain the cosmic shear B-mode on almost all scales.
On the other hand, the angular power spectrum of the curl mode does not correspond to the power of the Limber approximation on large scales, i.e., $\ell \lesssim 100$.
We can conclude that the totally coherent model works in the case of the cosmic shear B-mode.
Contrary to this, the totally coherent model is not reliable in the case of the CMB lensing curl mode at $\ell \lesssim 100$.

Here, we show the rough estimate for the signal-to-noise ratio in the case of the CMB lensing curl mode.
In the worst case, when the contribution from $\ell \lesssim 100$ on the signal-to-noise ratio is negligible, we find that $\epsilon_{v}$ decreases as $\epsilon_{v} \sim 1.8\times 10^{-5}$.
Although this value is most pessimistic constraint on the theoretical parameter of the texture by using the CMB lensing curl mode, it is comparable to the LSST case in the cosmic shear B-mode observation.
Therefore, the constraint on the theoretical parameter is at least $\epsilon_{v} \lesssim 1.8\times 10^{-5}$ by using the CMB lensing curl mode.

%============== Section ==============%
\section{Summary}\label{sec: SUM}
In this paper, we investigate weak lensing effects from non-topological textures accurately governed by the non-linear sigma model.
The phase transitions of the universe induce cosmological defects, e.g., monopoles, strings, or textures.
These defects imprint characteristic signatures on cosmological probes such as the CMB fluctuations or the large-scale structure.
We can give the constraint on cosmological defects from various observations.
Moreover, we can pull information indirectly about cosmological phase transitions which would have happened in the early universe.

In this paper, we examine weak lensing effects.
We can decompose weak lensing effects into two types of the signature by using the parity.
The parity-odd signal in weak lensing effects is induced from only the vector and tensor modes.
The dynamics of the non-topological texture is well described by the non-linear sigma model which induces not only scalar but also vector and tensor modes.
In order to estimate the weak lensing signal, we need to calculate the unequal-time power spectrum for the vector and tensor modes.
Throughout this paper, to proceed the numerical calculation, we restrict the totally coherent model for the texture which gives the unequal-time power spectrum is written by the separable form.
We leave to future work the consideration of any other models of the unequal-time power spectrum.

We present the CMB lensing curl-mode and cosmic shear B-mode from the non-topological texture with the large-$N$ limit.
In both observables, we newly find that the tensor mode dominates over the angular power spectrum of the curl and B modes.
We estimate the signal-to-noise ratio as a function of the theoretical parameter $\epsilon_{v}$.
The parameter $\epsilon_{v}$ represents the energy scale of the VEV.
In the current observations, the upper bound of $\epsilon_{v}$ is roughly obtained from the CMB anisotropies observed by the Planck as $\epsilon_{v} \lesssim 1.3\times 10^{-5}$ \cite{Ade:2013xla}.
Furthermore, the cosmic defects including the texture also induce the CMB spectral distortion \cite{Amin:2014ada}.
The CMB spectral distortion constrained by the COBE FIRAS \cite{Mather:1993ij} also imposes the upper bound as $\epsilon_{v} \lesssim 1.29\times 10^{-5}$, which is almost the same upper bound as the CMB anisotropies.
Note that, if we naively convert the tension of cosmic strings into the parameter $\epsilon_{v}$, $\epsilon_{v}$ for cosmic strings reads as $\epsilon_{v} \lesssim O(10^{-4})$ \cite{Ade:2013xla,2011JCAP...12..021U}.
The explicit bound depends on the kind of cosmic strings.

From our analysis, we find that the CMB lensing measurement by using the quadratic estimator without the instrumental noise would give an upper limit as $\epsilon_{v} \sim 2.7\times 10^{-6}$.
In the cosmic shear measurement, we give a relation between the signal-to-noise ratio and the survey design parameters.
From this result, improving the mean redshift is effective for studying the non-topological texture in the cosmic shear experiment.

%============= Acknowledgements =============%
\begin{acknowledgments}
This work was supported in part by a Grant-in-Aid for Japan Society for the Promotion of Science Research under Grants No.~14J00063 (S.S.) and No.~15J05029 (K.H.)
and a JSPS Grant-in-Aid for Scientific Research under Grant No. 24340048 (K.I.).
I also acknowledge the Kobayashi-Maskawa Institute for the Origin of Particles and the Universe, Nagoya University, for providing useful computing resources for conducting this research.
\end{acknowledgments}
%=============== Bibtex ===============%
\bibliography{ref}
\end{document}